\documentclass[twocolumn,english,pra,showpacs]{revtex4}
\usepackage[T1]{fontenc}
\usepackage[latin9]{inputenc}
\usepackage{amsmath}
\usepackage{amssymb}
\usepackage{graphicx}
\usepackage{esint}
\usepackage{amsfonts}
\usepackage{array}
\usepackage{multirow}
\usepackage{amstext}
\usepackage{CJK}
\usepackage{babel}
\usepackage{babel}
\usepackage{babel}
\usepackage{subfigure}
\usepackage{ulem}

\makeatletter
\@ifundefined{textcolor}{}
{
 \definecolor{BLACK}{gray}{0}
 \definecolor{WHITE}{gray}{1}
 \definecolor{RED}{rgb}{1,0,0}
 \definecolor{GREEN}{rgb}{0,1,0}
 \definecolor{BLUE}{rgb}{0,0,1}
 \definecolor{CYAN}{cmyk}{1,0,0,0}
 \definecolor{MAGENTA}{cmyk}{0,1,0,0}
 \definecolor{YELLOW}{cmyk}{0,0,1,0}
}
\makeatother

\usepackage{babel}
\begin{document}

\title{Collective effects of multi-scatterer on coherent propagation of
photon in a two dimensional network}

\author{D. Z. Xu$^{1}$, Yong Li$^{2}$, C. P. Sun$^{2}$, and Peng Zhang$^{3}$}

\email{pengzhang@ruc.edu.cn}

\selectlanguage{english}%

\affiliation{$^{1}$State Key Laboratory of Theoretical Physics, Institute of
Theoretical Physics, the Chinese Academy of Science and University of
the Chinese Academy of Sciences, Beijing 100190, China \\
 $^{2}$Beijing Computational Science Research Center, Beijing 100084,
China\\
 $^{3}$Department of Physics, Renmin University of China, Beijing
100872, China}

\begin{abstract}
We study the collective phenomenon in the scattering of a single photon
by one or two layers of two-level atoms. By modeling the photon dispersion
with a two-dimensional coupled cavity array (2D CCA), we analytically
derive the scattering probability of a single photon. We find
that the translational symmetry of the atomic distribution leads to
many important effects in the single-photon scattering. In the case
with one layer of atoms, the atomic collective Lamb shift
is related to the photonic density of states (DOS) of a 1D CCA,
rather than the photonic DOS of a 2D CCA. As a result, the photon is effectively \textit{not}
scattered by the atoms when the incident momentum of the photon takes
some special values. In the case with two layers of atoms, an inter-layer
effective coupling appears and induces an electromagnetic-induced-transparency-like
phenomenon. Our work provides a new scheme of analyzing photon
coherent transport in 2D and may help to understand the recent experiments
about the high energy photon scattering by the layered nuclei material.
\end{abstract}

\pacs{32.70.Jz, 32.80.Qk, 42.50.Nn}

\maketitle


\section{Introduction}

For the purpose of controlling the transport and the scattering of
a single photon in a full quantum fashion, much progress has been
made in confined atom-photon hybrid systems~\cite{Schoelkopf2004,Lukin2007}.
It is shown that the transport of a single photon can be coherently
manipulated by the interaction between the photon and the doped natural
or artificial atoms in low-dimensional quantum networks~\cite{SHFan2005,SHFanPRL2007,Zhou2008,LZhou2008}.
The corresponding physical implementations could be realized in several
different ways, such as: defected photonic crystals~\cite{Joannopoulos1995,Notomi2001}
and superconducting transmission line resonators~\cite{Nori2005,Liao2009}.
So far, many investigations have been done on one-dimensional (1D) atom-photon
hybrid system~\cite{Gong2008,supercavity,YChang}. For example,
people find that with a tunable two-level atom inside one of the cavities,
the transmission and reflection of a single photon can be well controlled
in 1D coupled cavity array (CCA) \cite{LZhou2008}. In order to fabricate
the integrated all-optical on-chip devices, it is also necessary to
study the control of the photon transport in two-dimensional (2D) structures
\cite{Altug2004,Majumdar2012}. Nevertheless, to our best
knowledge, this kind of investigation is still lack.

In this paper we begin to study the control of a
single-photon transport in a 2D CCA with atoms. The 2D CCA is a good
candidate for the 2D quantum networks. It has been experimentally
realized in defected photonic crystals \cite{Altug2004,Majumdar2012}.
The photon localization was observed in such a system with disorder
\cite{Topolancik2007}. Many authors proposed that the 2D CCA with
atoms can be used on quantum information \cite{Angelakis2008}
and the quantum simulation of many-body physics, such as the superfluid-Mott
transition \cite{Greentree2006,Hartmann2006} and quantum Hall effect
\cite{Bose2008,Girvin2010,Taylor2011}, \textit{etc}.

We investigate the scattering of a single photon in the
2D CCA, with two-level atoms which are periodically located in one
or two rows of the cavities. It is pointed out that, the research
for such a problem is not only helpful for the development of control
technique for photon transport in a 2D quantum network, but also important
for the investigations of collective effects of periodically located
atoms on the single-photon scattering. Since Dicke's initial work
on super-radiance \cite{Dicke1954}, many authors have studied the
influence of the collective behaviors of atoms on the spontaneous
radiation process \cite{Skribanowitz1973,Gross1982,Scully2009,Svidzindky2010,Skipetrov2011,Akkermans2013,Bonifacio1970,Jaynes1969,Savels2007}.
Most of these researches focused on the cases where the atoms
are randomly distributed \cite{Skipetrov2011,Akkermans2013}, or confined
in a region which is much smaller than the cube of the photonic wave
length \cite{Bonifacio1970}, or the systems with finite atoms \cite{Jaynes1969,Savels2007}.
Here we consider the \textit{scattering} process of a single photon,
and focus on the effects of infinite number of \textit{periodically}
distributed atoms, with the distance between two nearest atoms comparable
with the photonic wave length. Our research is also closely related
to the recent experiments of the single X-ray photon scattering on
the layered nuclear material, where the $^{57}$Fe nuclei
are periodically distributed \cite{Rohlsberger2010,Rohlsberger2012}.

For the 2D CCA system, we analytically solve the single-photon scattering
problem, and obtain a clear understanding of the physics behind the
collective effects. We find that when the atoms are located in one
row of the cavities in the 2D CCA, due to the translational symmetry
of the atomic distribution, the atomic collective energy shift induced
by the photon-atom interaction (i.e., collective Lamb shift) \cite{Friedberg1973}
is related to the photonic density of states (DOS) of a 1D CCA, rather than the photonic DOS of a 2D CCA. Furthermore, the collective Lamb shift diverges
when the incident momentum of the photon takes some special values. In
these cases the photon cannot be scattered by the atoms. When the
atoms are located in two rows of the cavities, as a result of the
periodical structure, only two atomic collective states are coupled
to the photonic states. Effective coupling between these two collective states, which is
described by the non-diagonal elements of the self-energy matrix, can be induced by
the photon-atom interaction. We can thus obtain two dressed
states with different effective energies, which are the eigen-states of the self-energy matrix. The maximum of the scattering
probability appears when the incident photon is resonant with one of these
two states. Similar as the atomic susceptibility in the system
with electromagnetic induced transparency (EIT), the scattering probability
has a double-peak behavior as a function of photon-atom detuning.

The rest of the paper is organized as follows. In Sec.~II we investigate
the single-photon scattering in a 2D CCA with one layer of atoms,
and discuss the collective shift of the atomic energy. In Sec.~III
we consider the case with two layers of atoms, and illustrate the
EIT-like behavior of the scattering probability. A brief conclusion is given 
in Sec. IV. Some details of our calculation are presented
in the appendix.

\section{single-photon scattering with one layer of atoms}

\subsection{System and Hamiltonian}

\begin{figure}
\includegraphics[width=8cm]{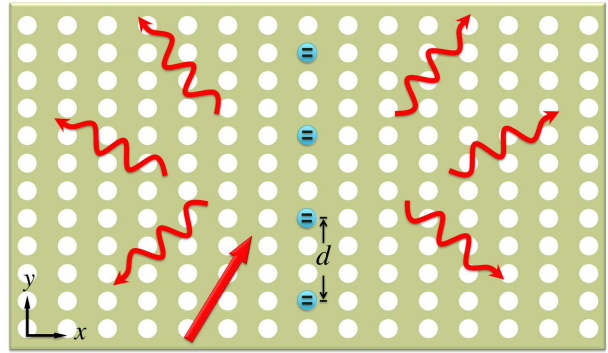}%
\caption{(Color online) The single-photon scattering by one layer of atoms
in a 2D CCA. The two-level atoms are confined in one row of the array
with period $d$ (in the figure we show the case with $d=3$). During
the scattering process, the incident photon (solid line) can be scattered
in several different outgoing directions (waved lines).}
\label{fig:1}
\end{figure}

We consider a 2D array of identical single-mode cavities as shown
in Fig.~\ref{fig:1}. We further assume that the photons can hop between
neighbor cavities. Then the cavity array is described by a tight-binding
model as
\begin{eqnarray}
H_{c} & = & \sum_{x,y=-\infty}^{+\infty}\omega_{c}a_{(x,y)}^{\dagger}a_{(x,y)}\notag\\
 &  & -\xi[a_{(x+1,y)}^{\dagger}a_{(x,y)}+a_{(x,y+1)}^{\dagger}a_{(x,y)}+h.c.].\label{hc}
\end{eqnarray}
Here $\omega_{c}$ is the frequency of the photons in the cavities,
$\xi$ is the hopping intensity or the inter-cavity coupling strength in both x and y directions, $a_{(x,y)}$ and $a_{(x,y)}^{\dagger}$ are the annihilation and creation
operators of the photon in the cavity at position $(x,y)$, respectively.
Here and after we set $\hbar=1$. When there is a single photon propagating
in the system, the eigenstate $|\vec{k}\rangle$ of $H_{c}$ takes
the form of 2D plane wave with momentum $\vec{k}=(k_{x},k_{y})$ as
\begin{equation}
|\vec{k}\rangle=\frac{1}{2\pi}\sum_{x,y=-\infty}^{+\infty}e^{i\left(k_{x}x+k_{y}y\right)}a_{(x,y)}^{\dagger}\left\vert \mathrm{vac}\right\rangle \label{|k>}
\end{equation}
with $\left\vert \mathrm{vac}\right\rangle $ the vacuum state of
all the cavities. The single-photon dispersive relation of the 2D
cavity array
\begin{equation}
\epsilon_{\vec{k}}\equiv\epsilon_{(k_{x},k_{y})}=\omega_{c}-2\xi(\cos k_{x}+\cos k_{y})\label{e_k}
\end{equation}
is naturally obtained by the stationary Schr$\ddot{\mathrm{o}}$dinger
equation $H_{c}|\vec{k}\rangle=\epsilon_{\vec{k}}|\vec{k}\rangle$.

To explore the scattering character of the incident photon on a collection
of identical atoms with geometrical configuration, we embed a two-level
atom in every $d$ cavities along the $y$-axis, and the $j$-th atom
is in the cavity at $(0,dj)$. The free Hamiltonian of these atoms is
\begin{equation}
H_{a}=\omega_{a}\sum_{j}\left\vert e\right\rangle _{j}\left\langle e\right\vert \label{Ha}
\end{equation}
with $\omega_{a}$ the energy-level spacing of the two-level atom
and $\left\vert e\right\rangle _{j}$ ($\left\vert g\right\rangle _{j}$)
the excited (ground) state of the $j$-th atom. The atom-photon coupling
is described by the Jaynes-Cummings Hamiltonian
\begin{equation}
V=\sum_{j}\Omega a_{\left(0,dj\right)}\left\vert e\right\rangle _{j}\left\langle g\right\vert +h.c. \label{V}
\end{equation}
with $\Omega$ the coupling strength.

The setup studied in this paper can be physically implemented by placing
the two-level atoms in the defected cavities of the 2D optical crystals
\cite{Altug2004}. Under the condition ($k_{x(y)}\sim\pi/2$),
the energy $\epsilon_{\vec{k}}$ becomes a linear function of the
momentum $\vec{k}$ as $\epsilon_{\vec{k}}\approx\omega_{c}+2\xi(k_{x}+k_{y})$.
Therefore, in this region our model can also characterize the scattering
of photons on the array of atoms in the free space, e.g., the scattering
process in the recent experiments with X-ray photon
scattered by the nuclei~\cite{Rohlsberger2010,Rohlsberger2012}.

\subsection{Single-photon scattering state and $T$-matrix}

Now we calculate the scattering probability of a single photon scattered
by the two-level atoms in our system. To this end, we first derive
the single-photon scattering state and the on-shell element of the
$T$-matrix in this subsection. With the help of these results, we
will obtain the single-photon scattering probability in the next subsection.

The scattering state $|\Psi^{\left(+\right)}\rangle$ is given by
the Lippman-Schwinger equation
\begin{equation}
|\Psi^{\left(+\right)}\rangle=|\vec{k}\rangle\left\vert \tilde{g}\right\rangle +\frac{1}{\epsilon_{\vec{k}}-(H_{a}+H_{c})+i0^{+}}V|\Psi^{\left(+\right)}\rangle\label{L-P_eq}
\end{equation}
with $\left\vert \tilde{g}\right\rangle \equiv\prod_{j}\left\vert g\right\rangle _{j}$
the collective ground state of all atoms. It is clear that in our
system the total excitation number $\sum_{x,y=-\infty}^{+\infty}a_{(x,y)}^{\dagger}a_{(x,y)}+\sum_{j}\left\vert e\right\rangle _{j}\left\langle e\right\vert $
is conserved. Thus, in the subspace with one excitation we can expand
the stationary eigenstate as
\begin{equation}
|\Psi^{\left(+\right)}\rangle=|\phi\rangle|\tilde{g}\rangle+\sum_{j}\beta_{j}\left\vert \mathrm{vac}\right\rangle |\tilde{e}_{j}\rangle.\label{SEform}
\end{equation}
Here $|\phi\rangle$ is a single-photon state of the cavity modes. The state
$\left\vert \tilde{e}\right\rangle _{j}$ is defined as $\left\vert \tilde{e}_{j}\right\rangle \equiv\left\vert e\right\rangle _{j}\otimes\prod_{l\neq j}\left\vert g\right\rangle _{l}$
and represents the state with only the $j$-th atom excited. It
follows from Eqs.~(\ref{L-P_eq}) and (\ref{SEform})
that $|\phi\rangle$ and $\beta_{j}$ satisfy
\begin{eqnarray}
|\phi\rangle & = & |\vec{k}\rangle+\sum_{j}\frac{\Omega^{\ast}\beta_{j}}{\epsilon_{\vec{k}}-H_{c}+i0^{+}}a_{(0,dj)}^{\dagger}\left\vert \mathrm{vac}\right\rangle ,\label{phi_la}\\
\beta_{j} & = & \frac{\Omega}{\epsilon_{\vec{k}}-\omega_{a}}\left\langle \mathrm{vac}\right\vert a_{(0,dj)}|\phi\rangle.\label{beta_la}
\end{eqnarray}

Eqs.~(\ref{phi_la}) and (\ref{beta_la}) can be analytically solved by
the following approach. First, we notice that our system is invariant
under the translation for $d$ cavities along the $y$-axis. Namely,
the total Hamiltonian $H_{a}+H_{c}+V$ of our system is commutative
with the translation operator $D$, which is defined as $Da_{(x,y)}D^{\dagger}=a_{(x,y+d)}$,
$D\left\vert \tilde{e}_{j}\right\rangle \langle\tilde{e}_{j}|D^{\dagger}=\left\vert \tilde{e}_{j+1}\right\rangle \langle\tilde{e}_{j+1}|$
and $D\left\vert \tilde{g}\right\rangle \langle\tilde{g}|D^{\dagger}=\left\vert \tilde{g}\right\rangle \langle\tilde{g}|$.
As a result of this symmetry, the scattering state $|\Psi^{\left(+\right)}\rangle$
in Eq.~(\ref{L-P_eq}) satisfies $D|\Psi^{\left(+\right)}\rangle=\exp(ik_{y}d)|\Psi^{\left(+\right)}\rangle$.
Therefore, we can conclude that the coefficient $\beta_{j}$ takes the
form
\begin{equation}
\beta_{j}=\beta e^{ik_{y}dj}.\label{bj}
\end{equation}

Second, substituting Eq.~(\ref{bj}) into Eqs.~(\ref{phi_la}, \ref{beta_la}),
we obtain the expression for the $j$-independent coefficient $\beta$:
\begin{equation}
\beta=\frac{\Omega}{2\pi}\frac{1}{\epsilon_{\vec{k}}-\omega_{a}-\Sigma(\vec{k})},\label{beta_k}
\end{equation}
where the self-energy $\Sigma(\vec{k})$ is given by
\begin{equation}
\Sigma(\vec{k})=\sum_{l=0}^{d-1}\Sigma_{l}(\vec{k}).\label{se}
\end{equation}
Here the function $\Sigma_{l}(\vec{k})$ is defined as
\begin{equation}
\Sigma_{l}(\vec{k})=\frac{\left\vert \Omega\right\vert ^{2}}{2\pi d}\int_{-\pi}^{\pi}dq_{x}\frac{1}{\epsilon_{\vec{k}}-
\epsilon_{\left[q_{x},p_{l}\left(k_{y}\right)\right]}+i0^{+}} \label{selfenergyl}
\end{equation}
with
\begin{equation}
p_{l}(k_{y})\equiv(k_{y}+\pi+\frac{2\pi|l|}{d})\mathrm{mod}[2\pi]-\pi.
\end{equation}
To obtain Eq.~$\left(\ref{selfenergyl}\right)$ we have also used the
formula
\begin{equation}
\sum_{j=-\infty}^{+\infty}e^{-i(q_{y}-k_{y})dj}=\frac{2\pi}{d}\sum_{l=0}^{d-1}\delta\left[q_{y}-p_{l}\left(k_{y}\right)\right],\label{sum}
\end{equation}
for $q_{y}$ $\in\left[-\pi,\pi\right]$. It is pointed out that,
in our system the self-energy $\Sigma(\vec{k})$ takes finite value
(except for some special momentums which will be discussed later)
and thus the renormalization technique is not required. This is due
to the fact that the single-photon energy $\epsilon_{\vec{k}}$ has
a finite upper limit $\omega_{c}+2\xi$.

Furthermore, we can treat the integral in $\Sigma_{l}(\vec{k})$
analytically and get the result
\begin{equation}
\Sigma_{l}(\vec{k})=\left\{ \begin{array}{ll}
-i\left\vert \Omega\right\vert ^{2}\left[2d\xi\left\vert \sqrt{1-A_{l}^{2}}\right\vert \right]^{-1}, & \left\vert A_{l}\right\vert <1, \\
\\
-\text{sign}(A_{l})\left\vert \Omega\right\vert ^{2}\left[2d\xi\sqrt{A_{l}^{2}-1}\right]^{-1}, & \left\vert A_{l}\right\vert >1,
\end{array}\right. \label{SeEnAn}
\end{equation}
with $A_{l}$ defined as
\begin{equation}
A_{l}\equiv\cos k_{x}+\cos k_{y}-\cos[p_{l}(k_{y})]\text{.}
\end{equation}
Substituting the result in Eq.~(\ref{SeEnAn}) into Eqs.~(\ref{SEform}-\ref{beta_la}),
we finally obtain the analytical expressions of the state $|\phi\rangle$,
the coefficient $\beta_{j}$, and the scattering state $|\Psi^{(+)}\rangle$.

With the analytical expression of the scattering state, we can calculate
the on-shell element $t(\vec{k}^{\prime}\leftarrow\vec{k})$ of the
$T$-matrix. According to the scattering theory \cite{Taylor1972},
$t(\vec{k}^{\prime}\leftarrow\vec{k})$ is defined as $t(\vec{k}^{\prime}\leftarrow\vec{k})=\langle\tilde{g}|\langle\vec{k}^{\prime}\mathbf{|}V|\Psi^{\left(+\right)}\rangle$.
The straightforward calculation yields
\begin{equation}
t(\vec{k}^{\prime}\leftarrow\vec{k})=u_{\mathrm{I}}(\vec{k})\sum_{l=0}^{d-1}
\delta\left(k_{y}^{\prime}-p_{l}(k_{y})\right),\label{t}
\end{equation}
where $\vec{k}^{\prime}=(k_{x}^{\prime},k_{y}^{\prime})$ and the
function $u_{\mathrm{I}}(\vec{k})$ is defined as
\begin{equation}
u_{\mathrm{I}}(\vec{k})=\frac{\left\vert \Omega\right\vert ^{2}}{2\pi d\left[\Delta-2\xi\left(\cos k_{x}+\cos k_{y}\right)-\Sigma(\vec{k})\right]}.\label{u}
\end{equation}
Here the photon-atom detuning $\Delta$ is defined as
\begin{equation}
\Delta=\omega_{c}-\omega_{a}.
\end{equation}
Due to the delta functions in Eq.~(\ref{t}), the $y$-component $k_{y}^{\prime}$
of the outgoing momentum can only take $d$ possible values. This
is also the result of the translation symmetry along the $y$-axis
in our system.

\subsection{Single-photon scattering probability}

Using the above results of the $T$-matrix element, we can calculate
the single-photon scattering probability. To this end, we consider
the scattering of a single-photon wave packet on the atoms. In the
scattering process, the incident wave packet of the photon can be
expressed as
\begin{equation}
|\Phi^{(in)}\rangle=\int d\vec{k}\phi^{\left(in\right)}(\vec{k})|\vec{k}\rangle\left\vert \tilde{g}\right\rangle .\label{phiin}
\end{equation}
Here $\phi^{\left(in\right)}(\vec{k})$ is single-photon wave function
in the momentum representation and satisfies $\int d\vec{k}|\phi^{\left(in\right)}(\vec{k})|^{2}=1$.
We further assume $\phi^{\left(in\right)}(\vec{k})$ sharply peaks
at a specific momentum $\vec{k}_{0}=(k_{0x},k_{0y})$. According to the
scattering theory, when the scattering process is completed, the single-photon
state can be expressed as
\begin{eqnarray}
|\Phi^{(out)}\rangle & = & \int d\vec{k}|\vec{k}\rangle\left\vert \tilde{g}\right\rangle \langle\tilde{g}|\langle\vec{k}|S|\Phi^{(in)}\rangle\label{sa}\\
 & \equiv & \int d\vec{k}\phi^{\left(out\right)}(\vec{k})|\vec{k}\rangle\left\vert \tilde{g}\right\rangle \label{sb}
\end{eqnarray}
in the interaction picture. Here the $S$-matrix satisfies
\begin{equation}
\langle\tilde{g}|\langle\vec{k}^{\prime}|S|\vec{k}\rangle\left\vert \tilde{g}\right\rangle =\delta(\vec{k}^{\prime}-\vec{k})-2\pi i\delta(\epsilon_{\vec{k}}-\epsilon_{\vec{k}^{\prime}})t(\vec{k}^{\prime}\leftarrow\vec{k}).\label{sc}
\end{equation}
Substituting Eqs. (\ref{t}, \ref{sc}) into Eqs. (\ref{sa}, \ref{sb}),
it is easy to find that after the scattering process, the incident
wave packet splits into $\left(2d-1\right)$ different ones (see Appendix).
Namely, the out-put wave function in Eq. (\ref{sb}) is given by

\begin{equation}
\phi^{\left(out\right)}(\vec{k})=\sum_{l=-\left(d-1\right)}^{d-1}\phi_{l}^{\left(out\right)}(\vec{k}).
\end{equation}
Here the $l$-th wave packet $\phi_{l}^{\left(out\right)}(\vec{k})$
sharply peaks at a momentum $\vec{k}_{l}=[k_{lx},p_{l}\left(k_{0y}\right)]$
with $\epsilon_{\vec{k}_{l}}=\epsilon_{\vec{k}_{0}}$ and
$\mathrm{sign}\left(k_{lx}\right)=\mathrm{sign}\left(l\right)$ (see
Appendix).

Then the probability for the photon being scattered to the $l$-th
($l\neq0$) outgoing momentum $\vec{k}_{l}$ is $P_{l}=\int d\vec{k}|\phi_{l}^{\left(out\right)}(\vec{k})|^{2}$.
As shown in the Appendix, $P_{l}$ can be expressed as
\begin{equation}
P_{l}=\frac{\left\vert u_{\mathrm{I}}(\vec{k}_{0})\right\vert ^{2}}{4\xi^{2}\left\vert \sin k_{0x}\sin k_{lx}\right\vert }.\label{pl}
\end{equation}
Therefore, for an incident photon with central momentum $\vec{k}_{0}$,
the scattering probability $R_{\mathrm{I}}(\vec{k}_{0})$ is
\begin{equation}
R_{\mathrm{I}}(\vec{k}_{0})=\sum_{l=-\left(d-1\right)}^{d-1}P_{l}=2\sum_{l=1}^{d-1}P_{l}+P_{0},\label{bigr}
\end{equation}
where we have used the fact $P_{l}=P_{-l}$.

\subsection{Behavior of the scattering probability}

\begin{figure}
\includegraphics[width=7cm]{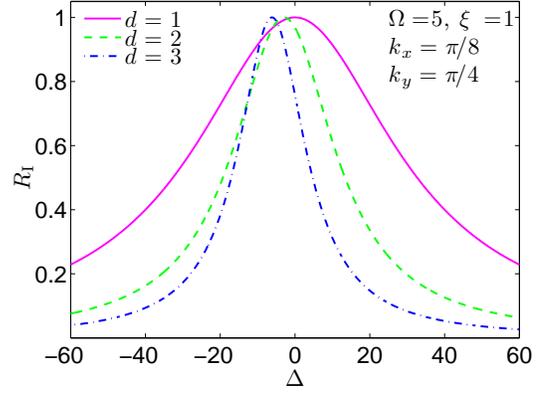} \caption{(Color online) The single-photon scattering probability $R_{\mathrm{I}}(\vec{k})$
as a function of photon-atom detuning $\Delta$ with $d=1$ (purple
solid line), 2 (green dash line) and 3 (blue dash-dotted line). Here
we choose $\vec{k}=(k_{x},k_{y})=\left(\pi/8,\pi/4\right)$ and $\Omega=5,\ \xi=1$.
The scattering probability $R_{\mathrm{I}}(\vec{k})$ has a Lorentz-shape
profiles and takes maximum value when $\Delta$ equals to
the collective Lamb shifts $\mathrm{Re}[\Sigma(\vec{k})]$.}

\label{fig:lorentz}
\end{figure}

Now we discuss the behavior of the scattering probability $R_{\mathrm{I}}(\vec{k})$
for an incident photon with central momentum $\vec{k}$. According
to Eqs.~(\ref{u}) and (\ref{pl}), it is clear that for fixed value
of $\vec{k}$, $R_{\mathrm{I}}(\vec{k})$ is a Lorentz function of
the photon-atom detuning $\Delta$, and takes the maximum value under
the condition $\Delta=2\xi\left(\cos k_{x}+\cos k_{y}\right)+\mathrm{Re}[\Sigma(\vec{k})]$.
In Fig.~\ref{fig:lorentz} we illustrate $R_{\mathrm{I}}(\vec{k})$
with different periods of atoms. The Lorentz-shape of the $R_{\mathrm{I}}(\vec{k})$
is clearly shown.

The single-peak behavior of $R_{\mathrm{I}}(\vec{k})$ can be explained
by the following simple picture. Owing to the periodical structure
of our system in the $y$-direction, in the single-photon scattering
process the incident state $|\vec{k}\rangle$ of the photon is coupled
to the atomic spin-wave state
\begin{equation}
|S_{k_{y}}\rangle\equiv\sum_{j}e^{ik_{y}dj}|\tilde{e}_{j}\rangle.
\end{equation}
As a result of this coupling, the effective energy of state $|S_{k_{y}}\rangle$
is shifted from the bare value $\omega_{a}$ to $\omega_{a}+\mathrm{Re}[\Sigma(\vec{k})]$.
Thus, $\mathrm{Re}[\Sigma(\vec{k})]$ can be considered as the collective
Lamb shift of the atomic state $|S_{k_{y}}\rangle$. The scattering
probability $R_{\mathrm{I}}(\vec{k})$ takes the maximum value when
the energy $\epsilon_{\vec{k}}$ of the incident photon equals to the shifted energy
$\omega_{a}+\mathrm{Re}[\Sigma(\vec{k})]$ of the spin-wave state. The cooperative effect of the atomic ensemble is reflected
by the atomic-density-dependence of this collective Lamb shift $\mathrm{Re}[\Sigma(\vec{k})]$
and the width $\mathrm{Im}[\Sigma(\vec{k})]$ of the peak of $R_{\mathrm{I}}(\vec{k})$
\cite{Ressayer1976,Gross1982,Scully2009,Svidzindky2010}.

\begin{figure*}
\includegraphics[width=12.5cm]{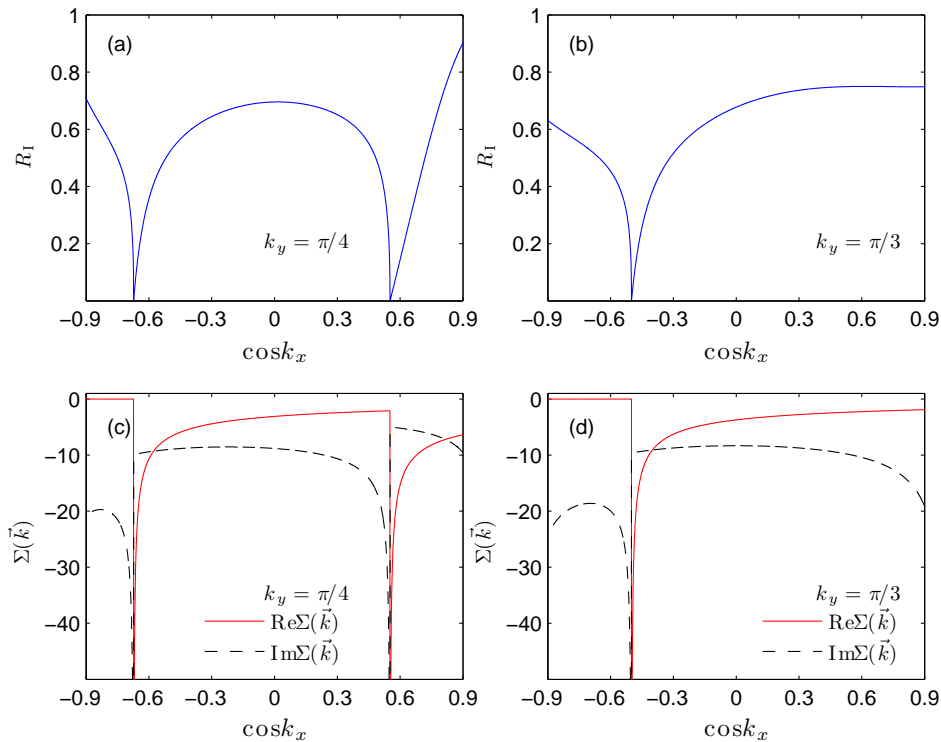} %
\caption{(Color online) (a) and (b): The single-photon scattering probability
$R_{\mathrm{I}}(\vec{k})$ as a function of $\cos k_{x}$ with $k_{y}=\pi/4$
(a) and $\pi/3$ (b). (c) and (d): The real part (red solid line)
and imaginary part (black dashed line) of self-energy $\Sigma(\vec{k})$
as functions of $\cos k_x$ with $k_{y}=\pi/4$ (c) and $\pi/3$ (d). Here, we also choose $\Omega=5,\ \xi=1,$ $\Delta=0$ and $d=3$. }

\label{fig:t}
\end{figure*}

Now we consider the effect of the translational symmetry of
the atomic distribution on the collective Lamb shift. Due to this
symmetry, the atomic spin-wave state $|S_{k_{y}}\rangle$ is only
coupled to the photonic states in the subspace ${\cal H}_{0}\oplus{\cal H}_{1}\oplus...\oplus{\cal H}_{d-1}$
during the scattering process. Here ${\cal H}_{l}$ is the subspace spanned
by the states $|\vec{k}_{l}\rangle\equiv|k_{lx},k_{ly}\rangle$, with
$k_{lx}\in(-\pi,\pi)$ and $k_{ly}$ taking a fixed value $p_{l}\left(k_{y}\right)$.
It is apparent that each space ${\cal H}_{l}$ is isomorphism to the
state space of a photon propagating in a 1D CCA,
rather than the one of a photon propagating in the 2D CCA. In the expression (\ref{se}) of the self-energy,
the term $\Sigma_{l}(\vec{k})$ is essentially contributed by the
coupling between $|S_{k_{y}}\rangle$ and the states in ${\cal H}_{l}$.
With straightforward calculation, we can re-write $\Sigma_{l}(\vec{k})$
defined in Eq.~(\ref{selfenergyl}) as
\begin{eqnarray}
\Sigma_{l}(\vec{k}) & = & \frac{\left\vert \Omega\right\vert ^{2}}{2\pi d}\int_{a_{l}}^{b_{l}}dE\frac{\rho(E)}{\epsilon_{\vec{k}}-E+i0^{+}}\notag\\
 & = & -i\frac{\left\vert \Omega\right\vert ^{2}}{2d}\rho(\epsilon_{\vec{k}})+\frac{\left\vert \Omega\right\vert ^{2}}{2\pi d}\mathrm{P}\int_{a_{l}}^{b_{l}}dx\frac{\rho(E)}{\epsilon_{\vec{k}}-E},\label{bb}
\end{eqnarray}
where P means the principle-value integral, $a_{l}=\epsilon_{\lbrack0,p_{l}(k_{y})]}$
and $b_{l}=\epsilon_{\lbrack\pi,p_{l}(k_{y})]}$ are the lower and
upper bounds of the energy $\epsilon_{\vec{k}_{l}}$ of the states in ${\cal H}_{l}$, respectively. In Eq.~(\ref{bb}) $\rho(x)$
is the density of states in ${\cal H}_{l}$ and can be expressed as
$\rho(x)=dk_{lx}/dE$, with $k_{lx}$ related to $E$ via the relation
$E=\epsilon_{\vec{k_{l}}}=\omega_{c}-2\xi[\cos k_{lx}+\cos p_l(k_{y})]$.
It is apparent that $\rho(x)$ is nothing but the density of states
of a single photon in a 1D CCA.

Usually, the self-energy $\Sigma_{l}(\vec{k})$ is convergent for a system with bounded energy spectrum. However, in the present problem, when the energy $\epsilon_{\vec{k}}$ of the incident photon is just at the boundaries of the energy spectrum of the states in
${\cal H}_{l}$, (i.e., the condition $\epsilon_{\vec{k}}=a_{l}$
or $\epsilon_{\vec{k}}=b_{l}$ is satisfied), the principle-value
integral in Eq.~(\ref{bb}) diverges. This observation is verified
by Eq.~(\ref{SeEnAn}) which shows that $\Sigma_{l}(\vec{k})=\infty$
when $|A_{l}|=\pm1$, i.e., $\epsilon_{\vec{k}}=a_{l}$ or $b_{l}$.
Furthermore, according to Eqs.~(\ref{u},\ref{pl},\ref{bigr}),
when the self-energy diverges we have $R_{\mathrm{I}}(\vec{k})=0$,
i.e., the photon is not scattered by the atoms. In Fig.~\ref{fig:t}
we plot the scattering probability $R_{\mathrm{I}}(\vec{k})$ and
the self-energy $\Sigma(\vec{k})$ as functions of $\cos k_{x}$ for
fixed values of $\Delta$ and $k_{y}$. It is clearly shown that $\Sigma(\vec{k})=\infty$
at the points where $R_{\mathrm{I}}(\vec{k})=0$.

In the end of this section, we would like to point out that, although $\Sigma_{l}(\vec{k})$
given in Eq.~(\ref{SeEnAn}) is proportional to the atomic density
$d^{-1}$, the total self-energy $\Sigma(\vec{k})$ is not a simple
linear function of $d^{-1}$, because $\Sigma(\vec{k})=\sum_{l=0}^{d-1}\Sigma_{l}(\vec{k})$
is the summation of $d$ terms. Therefore, the dependence of the collective
Lamb shift $\mathrm{Re}[\Sigma(\vec{k})]$ on atomic density is rather
complicated. For instance, in the cases shown in Fig.~2, the collective Lamb shift
increases when $d^{-1}$ is decreased from $1$ to $1/3$. We emphasize
that, the complicated relation between $\mathrm{Re}[\Sigma(\vec{k})]$
and $d^{-1}$ is caused by the expressions of $\Sigma(\vec{k})$ and
$\Sigma_{l}(\vec{k})$, and thus is essentially a result of the translational
symmetry of the atomic distribution in our system.

\section{Single-photon scattering with two layers of atoms}

\begin{figure}
\includegraphics[width=8cm]{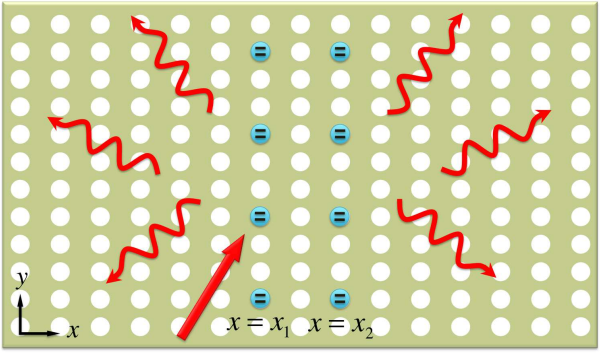} \caption{(Color online) The single-photon scattering by two layers of atoms
in a 2D CCA. In this case the atoms are confined in two layers at
$x=x_{1}$ and $x=x_{2}$, with the same period $d$. }
\label{fig:cavity2}
\end{figure}

In the above section we have studied the single-photon scattering
on one layer of atoms in a 2D cavity array. We show that the single-photon
scattering probability takes the maximum value when the incident photon
is resonant with the shifted atomic energy $\omega_{a}+\mathrm{Re}[\Sigma(\vec{k})]$,
and thus has a single peak as a function of the photon-atom detuning
$\Delta$. In this section, we consider the single-photon scattering
in the 2D cavity array with two layers of atoms located in the cavities
at $(x_{1},dj)$ and $(x_{2},dj)$ with $j=0,\pm1,\pm2,...$ (Fig.~\ref{fig:cavity2}).
We will show that as a function of $\Delta$, the scattering probability
has two peaks rather than a single one. The double-peak behavior is
due to the photon-induced effective coupling between atoms in different
layers, and can be considered as an EIT-like phenomenon.

In the presence of two layers of atoms, the atom-photon interaction
reads
\begin{equation}
V=\sum_{s=1,2}\sum_{j=-\infty}^{\infty}\Omega_{s}a_{(x_{s},dj)}\left\vert e\right\rangle _{j}^{(s)}\left\langle g\right\vert +h.c. \label{V2}
\end{equation}
with $\left\vert g(e)\right\rangle _{j}^{(s)}$ the ground (excited)
state of the $j$-th atom in the $s$-th layer. The scattering state
$|\Psi^{\left(+\right)}\rangle$ can be written as
\begin{equation}
|\Psi^{\left(+\right)}\rangle=|\phi\rangle|\tilde{g}\rangle+\sum_{s=1,2}\sum_{j=-\infty}^{\infty}\beta_{j}^{(s)}\left\vert \mathrm{vac}\right\rangle |\tilde{e}_{j}^{(s)}\rangle. \label{SEform2}
\end{equation}
Here $\left\vert \tilde{g}\right\rangle \equiv\prod_{j}\left\vert g\right\rangle _{j}^{(1)}\left\vert g\right\rangle _{j}^{(2)}$ is
the collective ground state of all atoms, and $|\tilde{e}_{j}^{(s)}\rangle$
($s=1,2$) defined as $|\tilde{e}_{j}^{(s)}\rangle\equiv\left\vert e\right\rangle _{j}^{(s)}\prod_{l\neq j}\left\vert g\right\rangle _{l}^{(s)}\prod_{n}\left\vert g\right\rangle _{n}^{(3-s)}$
denotes the state in which only the $j$-th atom in the $s$-th layer
is excited. Moreover, the translation invariance along the $y$-axis
leads to the result $\beta_{j}^{(s)}=\beta^{(s)}\exp[i(k_{x}x_{s}+k_{y}dj)]$.
Substituting this result into the Lippmann-Schwinger equation, we
find that the $j$-independent coefficients $\beta^{(1,2)}$ have
similar expressions with the parameter $\beta$ in Eq.~(\ref{beta_k}),
and can be written as
\begin{equation}
\left(\begin{array}{c}
\beta^{(1)}\\
\beta^{(2)}
\end{array}\right)=\frac{1}{2\pi\left[\omega_{\vec{k}}-\omega_{a}-\mathbf{\Sigma}(\vec{k})\right]}\left(\begin{array}{c}
\Omega_{1}\\
\Omega_{2}
\end{array}\right).\label{beta_equation}
\end{equation}
Here the self-energy $\mathbf{\Sigma}(\vec{k})$ is now a $2\times2$
matrix
\begin{equation}
\mathbf{\Sigma}(\vec{k})=\left[\begin{array}{cc}
\Sigma_{11} & \Sigma_{12}\\
\Sigma_{21} & \Sigma_{22}
\end{array}\right] \label{sig}
\end{equation}
with elements $\Sigma_{ij}$ ($i,j=1,2$) given by
\begin{equation}
\Sigma_{ij}=\frac{\Omega_{i}\Omega_{j}^{\ast}}{2\pi d}\sum_{l=0}^{d-1}\int_{-\pi}^{\pi}dq_{x}\frac{\exp[-i(k_{x}-q_{x})(x_{i}-x_{j})]}{\epsilon_{\vec{k}}-\epsilon_{[q_{x},p_{l}(k_{y})]}+i0^{+}}.\label{sij}
\end{equation}
It is clear that Eq.~($\ref{beta_equation}$) can be solved straightforwardly
and we have
\begin{equation}
\beta^{(s)}=\frac{\Omega_{s}}{2\pi\left(\Sigma_{+}-\Sigma_{-}\right)}
\left(\frac{\Sigma_{+}-J_{s}}{\Delta_{\vec{k}}-\Sigma_{+}}-\frac{\Sigma_{-}-J_{s}}
{\Delta_{\vec{k}}-\Sigma_{-}}\right) \label{beta2}
\end{equation}
for $s=1,2$. Here $\Delta_{\vec{k}}=\omega_{\vec{k}}-\omega_{a}$
and $J_{s}$ is defined as
\begin{equation}
J_{s}=\Sigma_{\left(3-s\right),\left(3-s\right)}-\frac{\Omega_{\left(3-s\right)}}{\Omega_{s}}\Sigma_{s,\left(3-s\right)}.
\end{equation}
In Eq.~(\ref{beta2}), $\Sigma_{\pm}(\vec{k})$ is the eigenvalue of
matrix $\mathbf{\Sigma}(\vec{k})$ and takes the form
\begin{equation}
\Sigma_{\pm}=\frac{1}{2}\left[\Sigma_{11}+\Sigma_{22}\pm\sqrt{\left(\Sigma_{11}-\Sigma_{22}\right)^{2}+4\Sigma_{12}\Sigma_{21}}\right].
\end{equation}
With these results, we can derive the expressions of the scattering
state $|\Psi^{\left(+\right)}\rangle$, and the on-shell element $t(\vec{k}^{\prime}\leftarrow\vec{k})$
of the $T$-matrix is
\begin{equation}
t(\vec{k}^{\prime}\leftarrow\vec{k})=u_{\mathrm{II}}(\vec{k})\sum_{l}\delta\left[k_{y}^{\prime}-p_{l}\left(k_{y}\right)\right].
\end{equation}
Now the coefficient function $u_{\mathrm{\mathrm{II}}}(\vec{k})$
is given by
\begin{equation}
u_{\mathrm{II}}(\vec{k})=\frac{1}{2\pi d}\sum_{s=1,2}e^{-i(k_{x}^{\prime}-k_{x})x_{s}}\Omega_{s}^{\ast}\beta^{(s)}.\label{u2}
\end{equation}

\begin{figure*}
\subfigure{ \includegraphics[width=6.5cm]{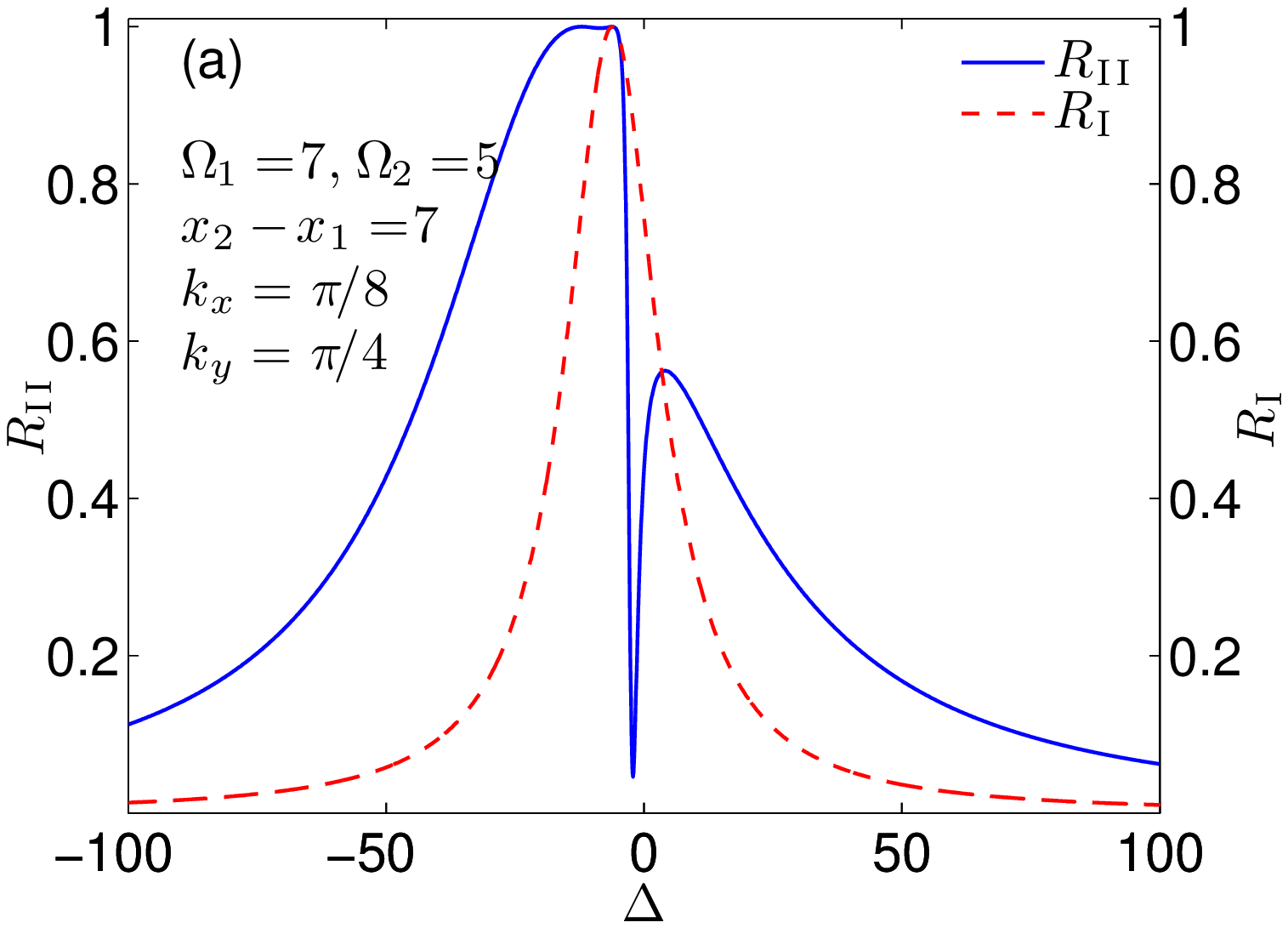}} \subfigure{ \includegraphics[width=6.5cm]{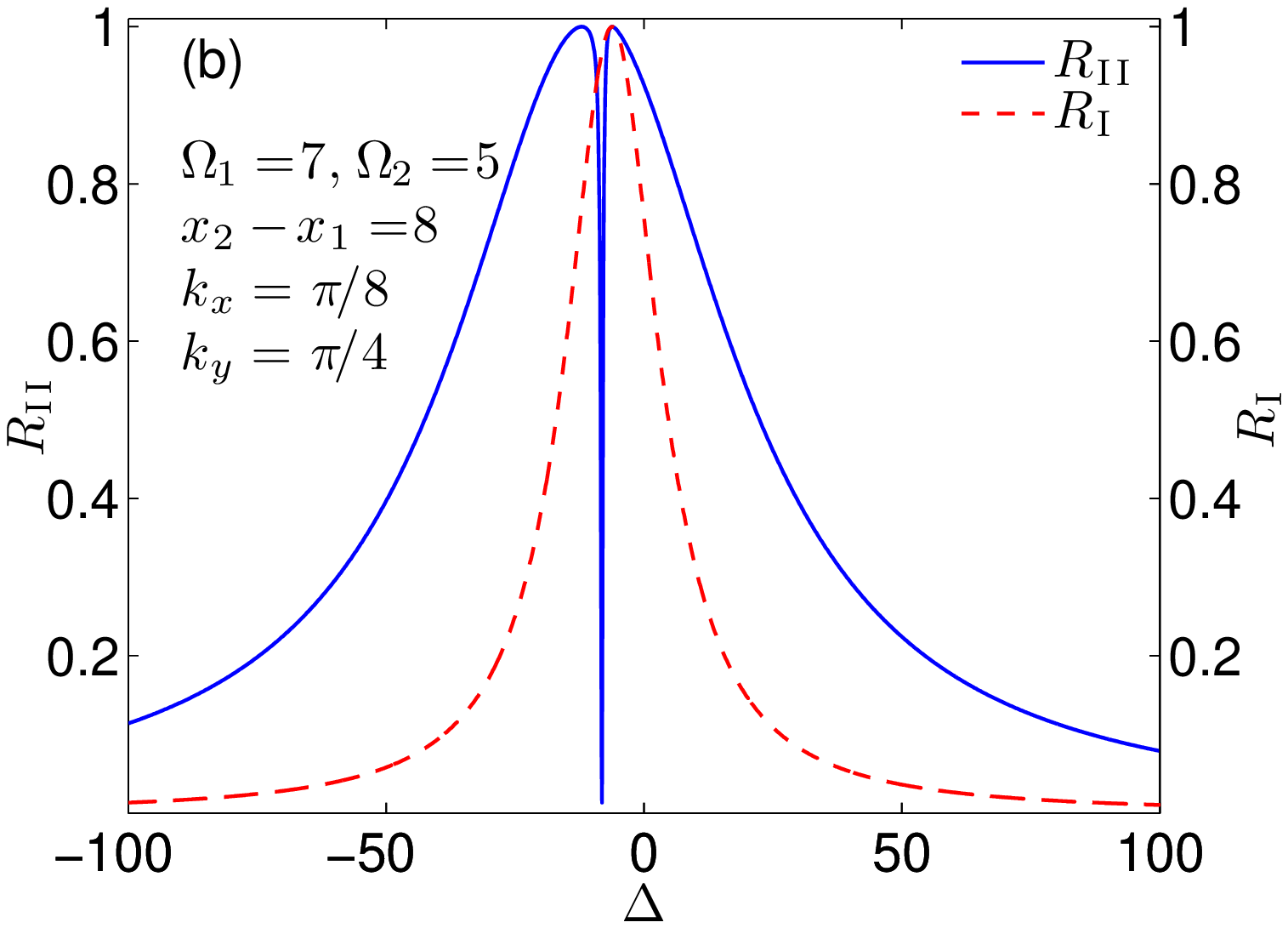}}
\subfigure{ \includegraphics[width=6.5cm]{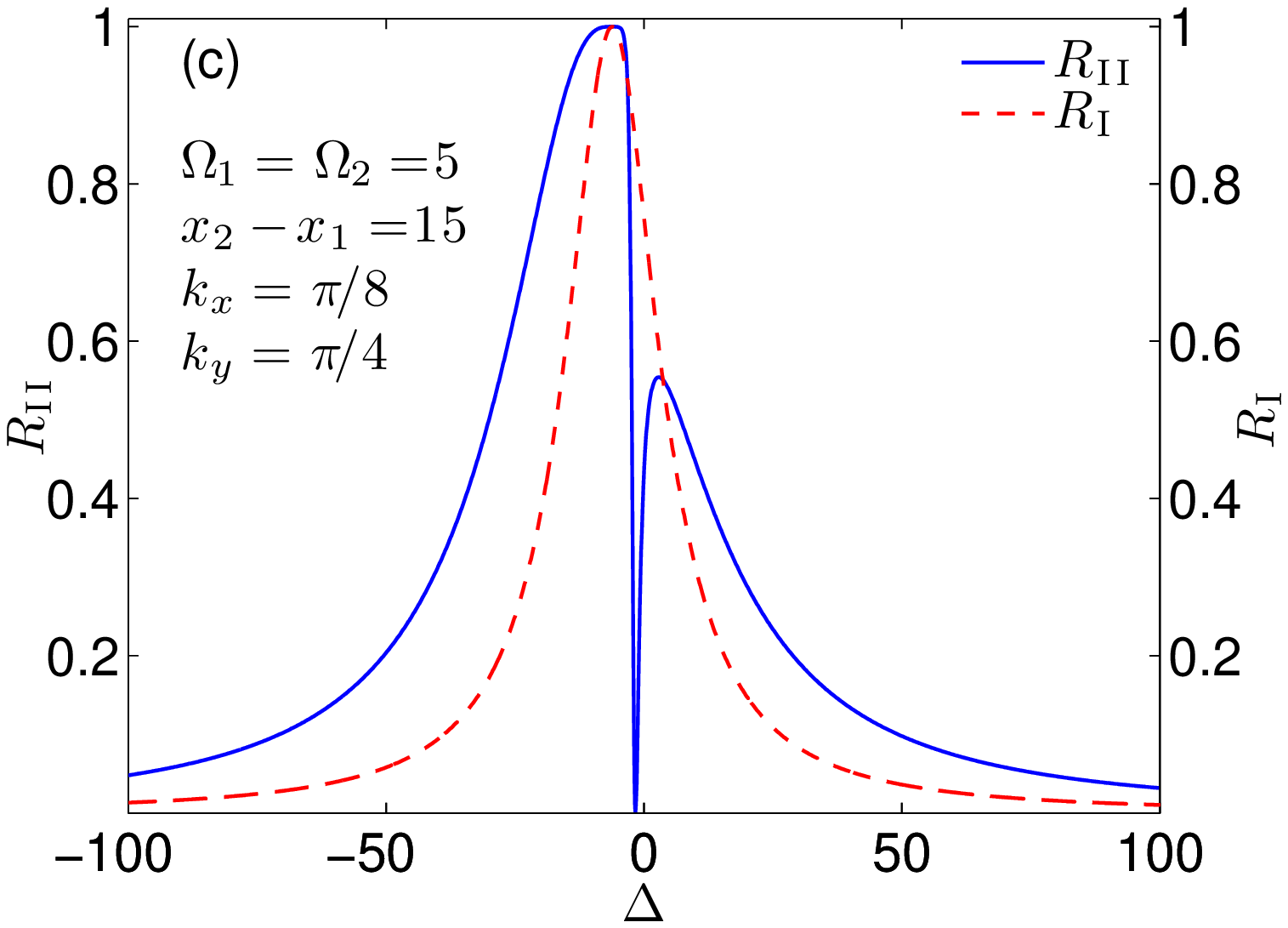}} \subfigure{ \includegraphics[width=6.5cm]{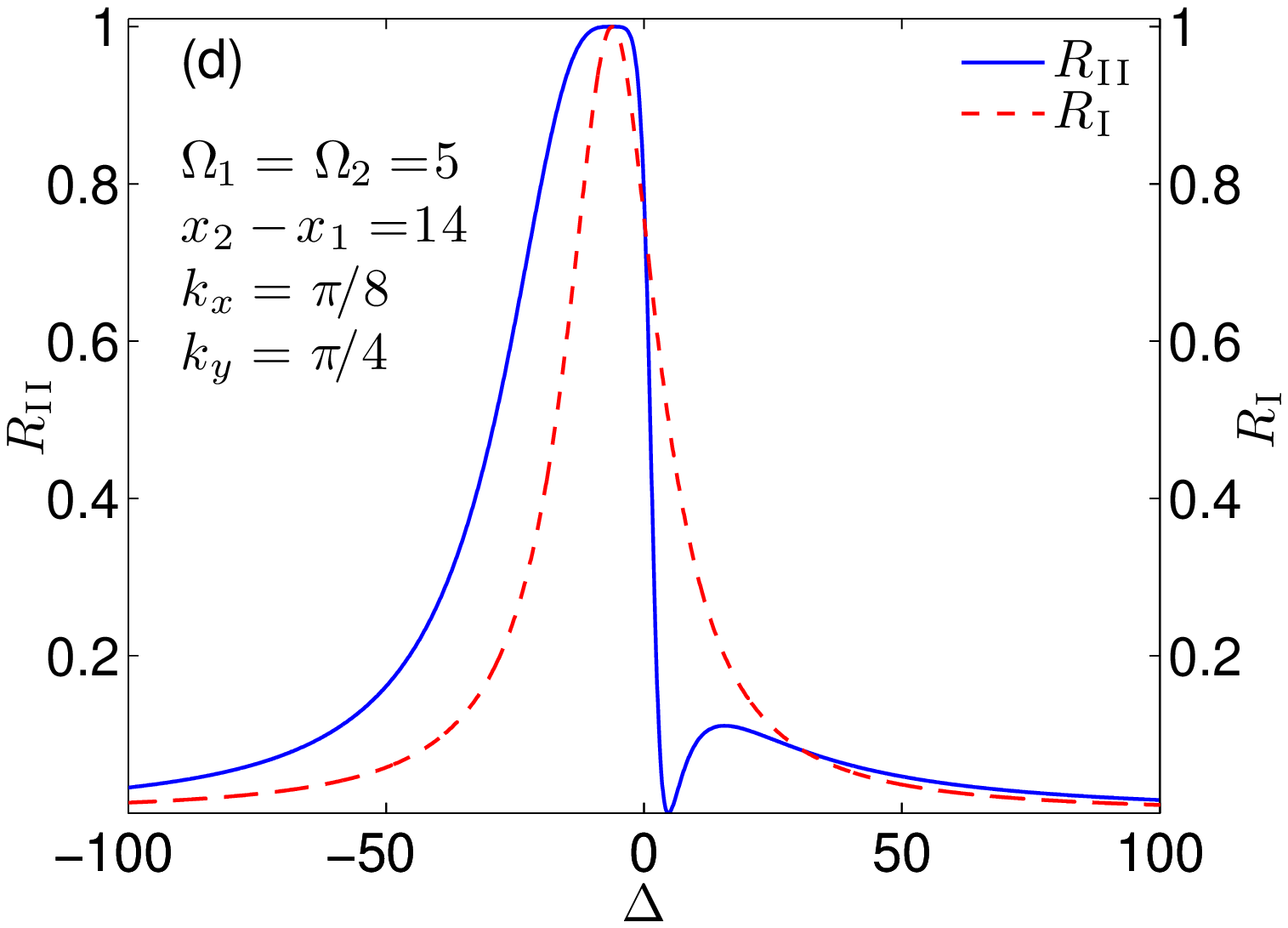}}
\subfigure{ \includegraphics[width=6.5cm]{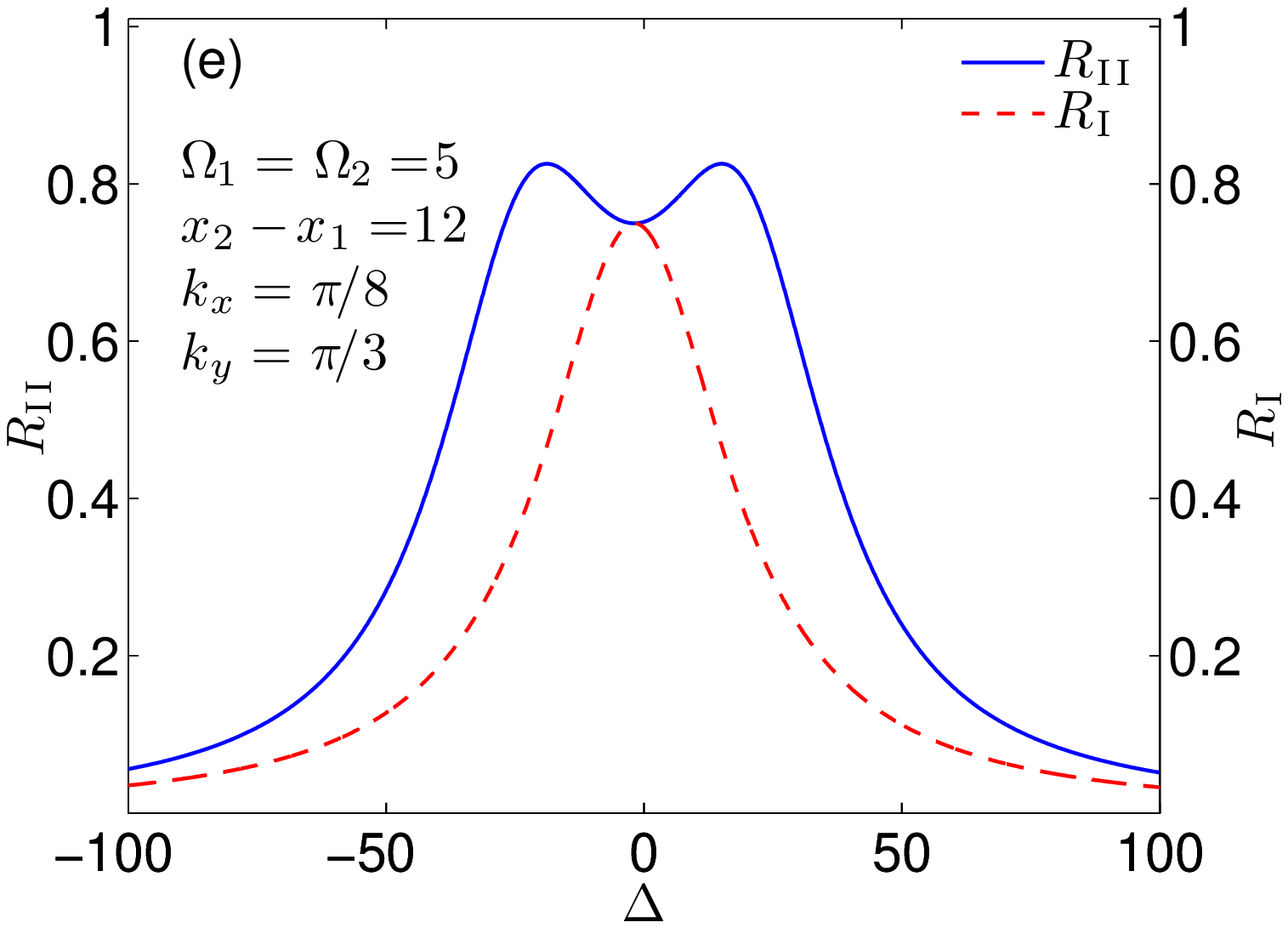}} \subfigure{ \includegraphics[width=6.5cm]{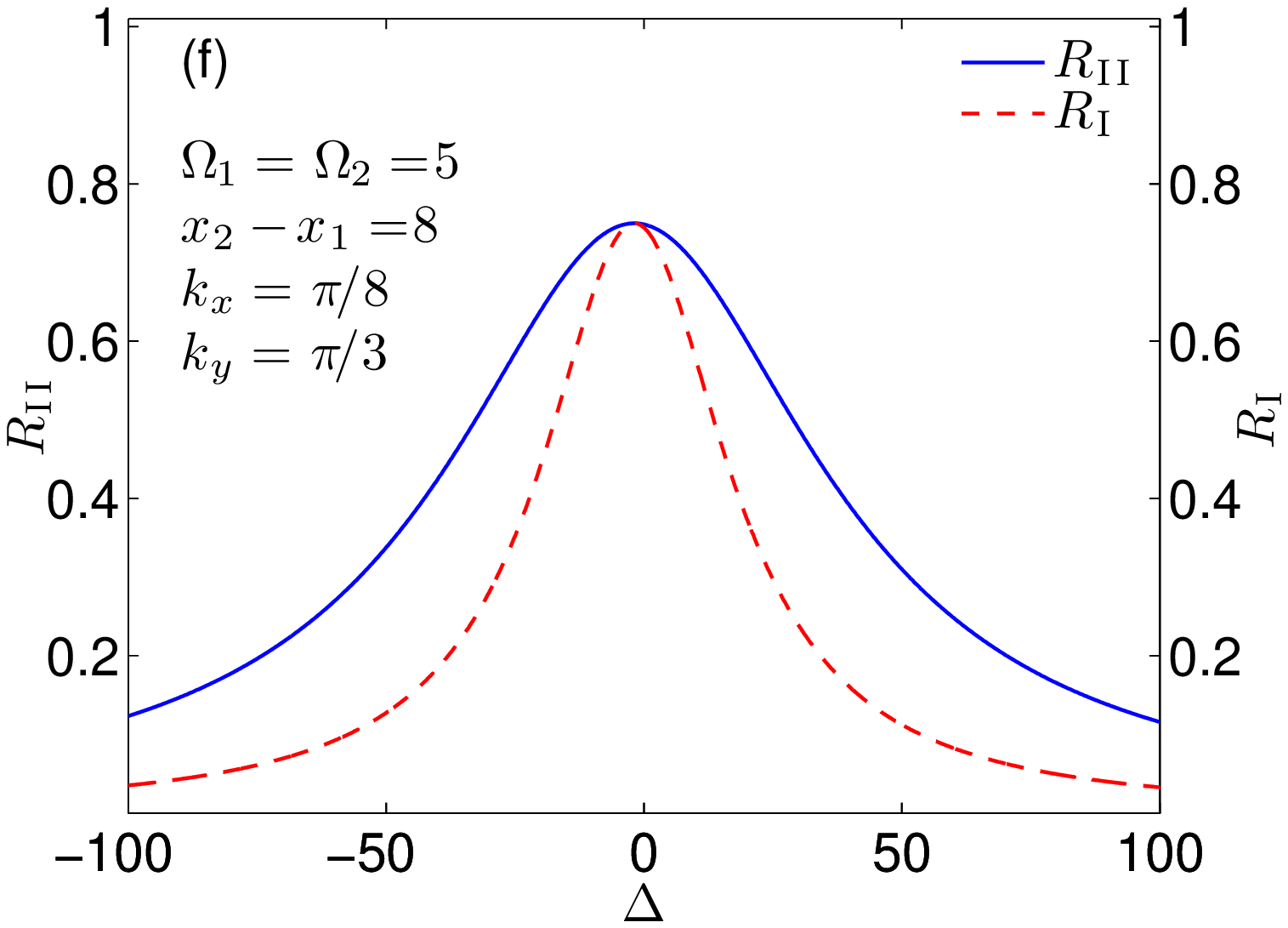}}
\caption{(Color online) The single-photon scattering probability $R_{\mathrm{II}}(\vec{k})$
for the two-layer case (solid blue line). The peak behavior of $R_{\mathrm{II}}(\vec{k})$
at $\Delta=\Delta_{\pm}$ is illustrated for the cases with $\Omega_{1}=7,\ \Omega_{2}=5$,
$\left(k_{x},k_{y}\right)=\left(\pi/8,\pi/4\right)$ and $x_{2}-x_{1}=7$
(a), $x_{2}-x_{1}=8$ (b) , as well as the cases with $\Omega_{1}=\Omega_{2}=5$
and $\left(k_{x},k_{y}\right)=\left(\pi/8,\pi/4\right)$, $x_{2}-x_{1}=15$
(c), $\left(k_{x},k_{y}\right)=\left(\pi/8,\pi/4\right)$, $x_{2}-x_{1}=14$
(d), $\left(k_{x},k_{y}\right)=\left(\pi/8,\pi/3\right)$, $x_{2}-x_{1}=12$
(e) and $\left(k_{x},k_{y}\right)=\left(\pi/8,\pi/3\right)$, $x_{2}-x_{1}=8$
(f). Here we choose $\xi=1$ and $d=3$. As a comparison, for each
parameter $(\xi,\ k_{x},\ k_{y},\Omega_{1},\Omega_{2})$ we also plot
the scattering probability $R_{\mathrm{I}}(\vec{k})$ for the corresponding
single-layer case with $\Omega=\Omega_{2}$ (dashed red line).}

\label{fig:DCS2b}
\end{figure*}

Similar to the above section, for an incident photon with central
momentum $\vec{k}$, the scattering probability $R_{\mathrm{II}}(\vec{k})$
can be expressed in terms of the $T$-matrix element, which is given
by
\begin{equation}
R_{\mathrm{II}}(\vec{k})=2\sum_{l=1}^{d}\frac{\left\vert u_{\mathrm{II}}(\vec{k})\right\vert ^{2}}{4\xi^{2}\left\vert \sin k_{x}\sin k_{lx}\right\vert } \label{r2}
\end{equation}
with $k_{lx}$ determined by the equation $\epsilon_{\vec{k}}=\epsilon_{\lbrack k_{lx},p_{l}(k_{y})]}$.

Now we investigate the behavior of the scattering probability $R_{\mathrm{II}}(\vec{k})$
with respect to the bare detuning $\Delta$. With Eqs.~(\ref{beta2})
and (\ref{u2}), we find that $R_{\mathrm{II}}(\vec{k})$ takes
local maximum values when the condition $\mathrm{Re}[\Delta_{\vec{k}}-\Sigma_{\pm}(\vec{k})]=0$
is satisfied. Namely, unlike the single-peak behavior shown in Eq.~(\ref{beta_k})
and Fig.~\ref{fig:lorentz} for the single-layer case, in the current
system $R_{\mathrm{II}}(\vec{k})$ has two peaks around the positions
\begin{equation}
\Delta_{\pm}\equiv\mathrm{Re}[\Sigma_{\pm}]+2\xi\left(\cos k_{x}+\cos k_{y}\right).
\end{equation}
Furthermore, in the two-layer case, the cooperative effect of the
atoms is reflected in the dependence of the collective Lamb shifts
$\mathrm{Re}[\Sigma_{\pm}(\vec{k})]$, the peak widths $\mathrm{Im}[\Sigma_{\pm}(\vec{k})]$
and the distance between the two peaks $\mathrm{Re}[\Sigma_{+}(\vec{k})-\Sigma_{-}(\vec{k})]$
on the atomic density $d^{-1}$ .

\begin{figure}
\includegraphics[width=8cm]{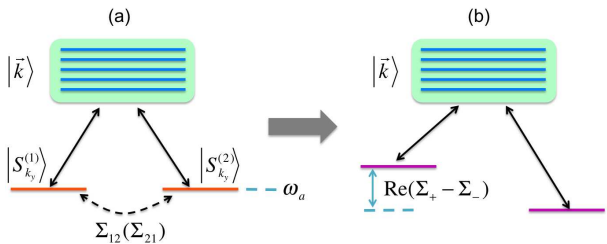} \caption{(a): In the two-layer case, the two quasi spin-wave states $|S_{k_{y}}^{(1)}\rangle$
and $|S_{k_{y}}^{(2)}\rangle$, which have the same bare energy $\omega_{a}$,
are coupled to the continuous spectrum of the photonic states. Due
to this atom-photon coupling, $|S_{k_{y}}^{(1)}\rangle$ and $|S_{k_{y}}^{(2)}\rangle$
are effectively coupled with each other by the non-diagonal terms
$\Sigma_{12}$ and $\Sigma_{21}$ of the self-energy matrix $\mathrm{\Sigma}(\vec{k})$.
(b): As a result of this effective coupling, $|S_{k_{y}}^{(1)}\rangle$
and $|S_{k_{y}}^{(2)}\rangle$ form two dressed states with different
effective energies $\omega_{a}+\Sigma_{\pm}$.}
\label{fig:effcouple}
\end{figure}

This observation is verified by our exact numerical calculation for
Eq.~(\ref{r2}). In Fig.~\ref{fig:DCS2b}, we illustrate $R_{\mathrm{II}}(\vec{k})$
for the two-layer case (with the comparisons to the one-layer case).
The double-peak behavior of $R_{\mathrm{II}}(\vec{k})$ at $\Delta=\Delta_{\pm}$
is clearly shown in Figs.~(\ref{fig:DCS2b}a-\ref{fig:DCS2b}d). It
is pointed out that, such a behavior essentially has the same physical
mechanism as that of the atomic susceptibility in an EIT system,
where the two internal states of the $\Lambda$-type atom are dressed
with the control laser beam, and form two dressed states with different
energies. Then the atomic susceptibility takes local maximum value
when the incident photon is resonant with one of these two states.
In our problem, the incident state $|\vec{k}\rangle$ of the photon
is coupled to two quasi spin-wave states
\begin{equation}
|S_{k_{y}}^{(1,2)}\rangle\equiv\sum_{j}e^{ik_{y}dj}|\tilde{e}_{j}^{(1,2)}\rangle
\end{equation}
with respect to the excitations of the atoms in the $1$st and $2$nd
layers, respectively. These two states have the same bare energy $\omega_{a}$.
Nevertheless, due to the atom-photon coupling, $|S_{k_{y}}^{(1)}\rangle$
and $|S_{k_{y}}^{(2)}\rangle$ are effectively coupled with each other
via the non-diagonal elements $\Sigma_{12}$ and $\Sigma_{21}$ of
the self-energy matrix $\mathbf{\Sigma}(\vec{k})$, and form two many-atom dressed
states. These two dressed states are no longer degenerate (Fig.~\ref{fig:effcouple}).
They have different effective energies $\omega_{a}+\Sigma_{\pm}$.
The single-photon scattering probability $R_{\mathrm{II}}(\vec{k})$
takes local maximum values when the incident photon is resonant with
one of these two dressed states. Therefore, the double-peak behavior of $R_{\mathrm{II}}(\vec{k})$
is an EIT-like phenomenon.

The above qualitative analysis provides a physical picture
for the EIT-like behavior of the single-photon scattering probability
$R_{\mathrm{II}}(\vec{k})$. Due to the complicated expression Eq.~(\ref{sij})
of the complex self-energy matrix $\mathbf{\Sigma}(\vec{k})$, in
different parameter regions the widths and heights of the two peaks
of $R_{\mathrm{II}}(\vec{k})$ can be quite different, as shown in
Fig.~5. In particular, the scattering probability is sensitive to the distance
between two atomic layers $x_{2}-x_{1}$, because $\Sigma_{12}$ and
$\Sigma_{21}$ are \textit{oscillating} functions of $x_{2}-x_{1}$.
When $x_{2}-x_{1}$ is tuned, the behavior of the scattering probability
can be significantly changed. Moreover, as shown in Figs.~(\ref{fig:DCS2b}e-\ref{fig:DCS2b}f), when the peak-widths are larger than the inter-peak distance, the two peaks gradually emerge into a single one.

\section{Conclusion}

In this paper, we study the single-photon scattering by one or two
layers of identical two-level atoms in a 2D CCA. We obtain the exact analytical
expression for the single-photon scattering probability, and explore
the related cooperative effects. In the case with one layer of atoms,
a Lorentz-type peak of the scattering probability implies the collective
Lamb shift. Due to the translational symmetry of the atomic
system, the collective Lamb shift is determined by the photonic DOS of a 1D CCA, rather than the photonic DOS of a 2D CCA. As a result, when the incident momentum takes some special finite values, the collective Lamb shift diverges and the photon is effectively not scattered by the atoms. In the two-layer case, the periodicity of our system renders only two collective atomic states to couple with the photon. An EIT-like double-peak
behavior appears when the effective coupling eliminates the degeneracy
between them. Our approach and results will be helpful
for the study of the 2D photonic quantum devices and X-ray quantum
optics from the perspective of scattering.

\begin{acknowledgments}
This work is supported by the National Natural Science Foundation of China
under Grants No. 11121403, No. 10935010, No. 11222430, No. 11074305,
No. 11074261, No. 11174027 and the National Basic
Research Program of China under Grant No.~2012CB922104. PZ would also like to thank the NCET Program for support.
\end{acknowledgments}

\section*{Appendix: single-photon outgoing wave packets and probability $P_{l}$}

In this appendix we derive the outgoing wave function of a single photon
scattered by the atoms, and prove Eq.~(\ref{pl}) for the probability
$P_{l}$. Substituting Eqs.~(\ref{t},\ref{sc}) into Eqs.~(\ref{sa},\ref{sb}), it is easy to find that after the scattering process, the incident wave packet splits into $\left(2d-1\right)$ different ones. Namely, the out-going wave function in Eq.~(\ref{sb}) is given by
\begin{equation}
\phi^{\left(out\right)}(\vec{k})=\sum_{l=-d}^{d}\phi_{l}^{\left(out\right)}(\vec{k}).
\end{equation}
Here the function $\phi_{l}^{\left(out\right)}(\vec{k})$ is defined
as
\begin{equation}
\phi_{0}^{\left(out\right)}(\vec{k})=\frac{u(\vec{k})\phi^{\left(in\right)}(\vec{k})}{2\xi\left\vert \sin k_{x}\right\vert },\label{aa}
\end{equation}
for $l=0$ and
\begin{equation}
\phi_{l}^{\left(out\right)}(\vec{k})=\frac{u[\vec{f}_{l}(\vec{k})]\phi^{\left(in\right)}[\vec{f}_{l}(\vec{k})]}{2\xi\left\vert \sin f_{lx}(\vec{k})\right\vert }\delta_{\mathrm{sign}\left(k_{x}\right),\mathrm{sign}\left(l\right)}\label{cc}
\end{equation}
with the function $\vec{f}_{l}(\vec{k})=\left[f_{lx}(\vec{k}),f_{ly}(\vec{k})\right]$
defined as
\begin{eqnarray}
f_{lx}\left(\vec{k}\right) & = & \arccos\left\{ \cos k_{x}+\cos k_{y}-\cos\left[f_{ly}\left(\vec{k}\right)\right]\right\} ,\notag\\
\\
f_{ly}\left(\vec{k}\right) & = & (k_{y}+\pi+\frac{2\pi|l|}{d})\mathrm{mod}[2\pi]-\pi,
\end{eqnarray}
and the Kronecker symbol $\delta_{i,j}$ satisfies $\delta_{i,j}=1$
for $i=j$ and $\delta_{i,j}=0$ for $i\neq j$. In Eqs.~(\ref{aa})
and (\ref{cc}), we have $u(\vec{k})=u_{\mathrm{I}}(\vec{k})$ for
the case with one layer of atoms and $u(\vec{k})=u_{\mathrm{II}}(\vec{k})$
for the two-layer case. Then it is easy to see that, when the incident
wave packet $\phi^{\left(in\right)}(\vec{k})$ sharply peaks at a
specific momentum $\vec{k}_{0}=(k_{0x},k_{0y})$, the out-put wave
packet $\phi_{l}^{\left(out\right)}(\vec{k})$ sharply peaks at the
momentum $\vec{k}_{l}=[k_{lx},p_{l}\left(k_{0y}\right)]$ which satisfies
$\epsilon_{\vec{k}_{l}}=\epsilon_{\vec{k}_{0}}$ and $\mathrm{sign}\left(k_{lx}\right)=\mathrm{sign}\left(l\right)$.

Now we calculate the probability $P_{l}=\int|\phi_{l}^{\left(out\right)}(\vec{k})|^{2}d\vec{k}$
for the cases with $l\neq0$. Apparently, we have
\begin{equation}
P_{l}=\int\left\vert \frac{u[\vec{f}_{l}\left(\vec{k}\right)]\phi^{\left(in\right)}[\vec{f}_{l}\left(\vec{k}\right)]}{2\xi\left\vert \sin f_{lx}\left(\vec{k}\right)\right\vert }\delta_{\mathrm{sign}\left(k_{x}\right),\mathrm{sign}\left(l\right)}\right\vert ^{2}d\vec{k}.
\end{equation}
We define $g_{lx}=f_{lx}(\vec{k})$ and $g_{ly}=f_{ly}(\vec{k})$.
Then using the fact that $\phi^{\left(in\right)}(\vec{k})$ sharply
peaks at a specific momentum $\vec{k}_{0}$ and the relations
\begin{equation}
dk_{x}=\left\vert \frac{\sin g_{lx}}{\sin k_{x}}\right\vert dg_{lx},dk_{y}=dk_{ly},
\end{equation}
we have
\begin{eqnarray}
P_{l} & \approx & \int\left\vert \frac{u[\vec{f}_{l}(\vec{k})]\phi^{\left(in\right)}[\vec{f}_{l}(\vec{k})]}{2\xi\left\vert \sin f_{lx}(\vec{k})\right\vert }\right\vert ^{2}d\vec{k}\notag\\
 & \approx & \frac{\left\vert u(\vec{k}_{0})\right\vert ^{2}}{4\xi^{2}\left\vert \sin k_{0x}\sin k_{lx}\right\vert }\int\left\vert \phi^{\left(in\right)}(g_{lx},g_{ly})\right\vert ^{2}dg_{lx}dg_{ly}\notag\\
 & = & \frac{\left\vert u(\vec{k}_{0})\right\vert ^{2}}{4\xi^{2}\left\vert \sin k_{0x}\sin k_{lx}\right\vert }.
\end{eqnarray}
Then we have proved Eq.~(\ref{pl}).

\end{document}